\documentstyle[11pt,psfig]{article}

\small
\begin{document}
\newcommand{\be}{\begin{equation}}
\newcommand{\ee}{\end{equation}}

\begin{titlepage}
\title{
{\bf INFLUENCE OF RESCATTERING ON THE SPECTRA OF STRANGE PARTICLES} }
\author{ {\bf C. David}\thanks {email : "DAVID@nanhp2.in2p3.fr"}
{\bf ,~C. Hartnack}\thanks {email : "HARTNACK@nanhp2.in2p3.fr"}
{\bf ,~M. Kerveno}{\bf ,~J.-Ch. Le Pallec}
{ \bf and J. Aichelin}\thanks {email : "AICHELIN@nanvs2.in2p3.fr"}
\\
{\normalsize
SUBATECH}\\
{\normalsize
Universit\'e de Nantes, Ecole des Mines, IN2P3/CNRS }\\
{\normalsize 4 Rue Alfred Kastler Nantes, F-44070, France   }
}
\date{}
\maketitle
\begin{abstract}
Applying a new method of rescattering which is based on the neural network
technique we study the influence of rescattering on the spectra of 
strange particles produced in heavy ion reactions. In contradistinction
to former approaches the rescattering is done explicitly and not in a 
perturbative fashion. We present a comparison of our calculations for the
system Ni(1.93AGeV)+Ni with recent data of the FOPI 
collaboration. We find that even for this small system rescattering changes 
the observables considerably but does not  invalidate the role of the
kaons as a messenger from the high density zone. We cannot confirm
the conjecture that the kaon flow can be of use for the determination
of the optical potential of the kaon. 
\end{abstract}


\vskip 1cm
SUBATECH 96-09 \hfill{November 1996}

\vskip 1cm
{\it Submitted to Nucl.Phys.A }
\end{titlepage}

\newpage
\section{Introduction}
The production of kaons in heavy ion collisions is presently one of the
most challenging topics in nuclear physics. At beam energies below or
close to the threshold (in NN collisions) of $E_{beam} = 1.583$ GeV we
observe a strong enhancement of the kaon production as compared to the
extrapolation of pp collisions. Detailed investigations have shown that
there most of the kaons are created in two step processes via an
intermediate $\Delta$ or $\pi$ and are produced at a density well above
the nuclear matter density \cite{aik,har}. This triggered the conjecture that kaons may be
of use as a messenger of the high density zone. In this high 
density zone the compressional energy may be high (depending  on the
nuclear equation of state) and therefore, due to energy conservation,
the available kinetic energy may be reduced. Hence we have to expect
a lowering of the production rate. However, at this high density
also the properties of the kaon may have changed. Due to the
presence of strong scalar and vector fields the threshold for
kaon production may be lowered leading to an enhancement of the kaon yield.
Calculations have shown that the fields cause only a small correction
for the $K^+$ but for the $K^-$ the effect may be large due to g - parity
\cite{sch}.

In infinite matter these phenomena can be studied using chiral
Lagrangians. The higher the density and the temperature 
the more difficult it is, however,
to solve the perturbation expansion of this Lagrangian. In addition
a verification of the results in experiments is only possible with heavy
ions. These heavy ion reactions are unfortunately far away
from thermal equilibrium and finite size effects are dominant. The
presently only possibility to face this problem are sophisticated
numerical programs which simulate the time evolution of the
reaction. These programs may reveal the influence of different conjectures
on the meson production and meson properties on the observable spectra.
 
Due to the smallness of the cross section the kaon production has
been calculated up to now only perturbatively: for every collision i with 
$\sqrt{s} > \sqrt{s}_{threshold}$ one has registered the production
probability to produce a kaon $P_i = \sigma(NN\rightarrow K)/\sigma
(NN\rightarrow X)$. The total kaon cross section is then given by
$\sigma_{K} = \sigma_{reac}\sum_i P_i$. This perturbative approach
does allow to predict the cross section for $K^+$ which due to its
$\bar s$ quark cannot be reabsorbed. The small change on the effective
$K^+$ yield in isospin asymmetric systems due to $K^+ + n \rightarrow
K^0 + p$ has been neglected. For generating the
spectra one assumes that the $N\Lambda (\Sigma)K$ system decays 
according to phase 
space in its center of mass system. This assumption is a first order 
approximation to existing data. About deviations conflicting results 
are reported\cite{hog,lag}. Applying Monte Carlo procedures one can perform
the disintegration of a single $N\Lambda (\Sigma)K$ system many times where
each disintegration yields a precise value of the momenta of all particles.
This allows a very effective calculation of differential kaon cross section.

In some of these simulation programs an approximative rescattering
procedure has been added which goes back to Randrup \cite{ran}.
There, after the determination of the momentum of
the kaon, one checks the path the kaon has to travel through
nuclear matter if the nucleus rest in the configuration given at the
moment of the creation of the kaon. The length of the path is then 
divided by the mean free
path of the kaon in nuclear matter to obtain the average collision number.
The number of collisions the kaon is supposed to suffer is given
by a Poisson distribution around this mean value.The momentum distribution of
the nucleons is taken from the measured proton spectrum. Furthermore, it is
assumed that the $KN$ cross section is purely elastic and the angular
distribution of the cross section is isotropic. Any feedback of the kaon
production on the system is neglected, i.e. all non strange particles move as if
there has been no kaon produced.
An improved version of this approximative rescattering procedure has been recently
proposed by Fang et al. \cite{fa}. Instead of calculating the
path of the kaons in matter using the density distribution as given at the instant 
of the creation of the kaon Fang et al. follow the trajectory of the
kaon, assuming that in between collisions the kaons move on straight
lines. Collisions with the nucleons change only the momentum of the
kaon whereas the nucleons move as if no collisions had appeared. The
cross section of KN scattering is taken as isotropic. Thus as
in the aforementioned approaches there is no 
feedback of the kaon creation on the nuclear system.

In the last years the experiments on meson production in heavy ion
collisions reached a precision which makes it necessary to minimize
the systematic errors due to approximations in the simulation programs.
One of the major sources of the systematic error is the approximate
treatment of rescattering. In addition the use of an unrealistic isotropic 
cross section does not allow to address a conjecture which
has been advanced recently: Does the kaon flow depend crucially on
the magnitude of the vector and the scalar potential and how compares
the kaon flow with that of the baryons? If kaons rescatter with baryons
which possess directed flow, this directed flow may be communicated to then
and shows up in the final kaon spectra. 

The production of a kaon is a rare event. Even at central collisions
of Ni+Ni at 1.93 A GeV, i.e. well above the threshold, a kaon is produced
only in 1 event out of one thousand. Therefore it is not easy to gain sufficient
statistics if one would like to follow the time evolution of the kaons
in the system. It is the purpose of this article to introduce a new method
to achieve this goal. In the next chapter we introduce this new
method, in chapter 3 we study the influence of rescattering on the
observables taking as an example the recent data obtained by the FOPI
detector.

\section{Neural Network for the Rescattering of Kaons}

Since the kaon production is a rare event in heavy ion collisions below
or close to the threshold we have to enhance artificially
its production rate if we want to avoid simulations which are useless
for studying kaon properties. This can be achieved by introducing
an artificial kaon production enhancement factor which assures
that on the average in about 30\% of the heavy ion collisions 1 kaon is produced.
Due to the fluctuation around the mean value we see in 7\% of the
reactions the production of two kaons, a quantity which we consider as 
tolerable.

This enhancement factor is a necessary but not a sufficient
condition to obtain a reasonable kaon yield in a short amount of time.
A second condition is that the treatment of the rescattering is done
fast and effectively despite of the rather complicated structure of the
elementary $KN$ cross section. This second condition is achieved with
a modern mathematical tool, called neural network. 

The basic problem in simulating a cross section by a Monte Carlo procedure
is to assure that after many simulations the particles are scattered with the desired differential 
cross section,i.e. that the distribution P(x) of scattering angles corresponds 
to the desired one. In one dimension and in the case that the integral
over the distribution can be inverted this can be achieved with elementary
methods.
\be
         y_i = \int_{lower\  bound}^{x_i} P(x) dx 
\ee
where $y_i$ is a random number uniformly distributed in [0,1]. The $x_i$'s
are distributed like $P(x)$ if $P(x)$ is normalized . Usually the cross section
is not given in an analytical form but by experimental data points and
the angular distribution varies as the energy increases. Then one
approaches usually the problem by multi dimensional fits. Multiparameter
fits are normally difficult to invert and in addition the inversion
poses problems due to the finite accuracy of the computer, a problem
one suffers already for pp collisions.

The neural network takes a different approach. First one has to 
fit the differential cross sections according to the data by polynomials
as follows
\be
{\frac{d\sigma}{d\Omega}=\sum_{i=0}^n a_i(\sqrt{s})(cos\theta)^i}
\ee
where $a_i(\sqrt{s})$ are the polynomial coefficients and $n$ the degree of the
polynomial. Then we calculate $X$ defined as
\be
{X(\sqrt{s},cos\theta^i)=\frac{\int_{-1}^{cos\theta^i}\frac{d\sigma}{d\Omega}
(\sqrt{s},cos\theta')d(cos\theta')}
{\int_{-1}^1\frac{d\sigma}{d\Omega}(\sqrt{s},cos\theta')d(cos\theta')}}
\ee

$X$ is calculated for many values of $cos\theta^i$ chosen between -1 et 1 at
the points of $\sqrt{s}$ where data exist. We see that $X$ is distributed 
between 0 and 1.

In the second step $X(\sqrt{s},cos\theta^i)$
and $\sqrt{s}$ are presented as input to a neural network
as displayed in fig.1. As an output variable $x_{out}^i$ we would like to have the
scattering angle $\cos{\theta}^i$. With help of the many values of X prepared as
above we train the network, i.e. we minimize the standard deviation
\be
\sum (\cos{\theta}^i - x_{out}^i)^2 
\ee 
by varying the free parameters of the network. For the detailed description
of the free parameters of the neural network and the method to achieve
this goal we refer to reference \cite{dav}. After this procedure one calls
the network "trained". The quality of the training can be viewed from fig. 2
where we display the response of the trained network. For three different
momenta we display $\cos{\theta}^i$ given as input into the network 
in form of eq. 3 and
the $\cos{\theta}^i$ estimated by the neural network, i.e. the output
value of the network. For a perfect reproduction we expect a straight line.
As seen, the deviations are of minor importance.

Now the network is prepared to serve the purpose. For each collision
we present to the network the center of mass energy ${\sqrt{s}}$ and a
random number and the network responds with a scattering angle. The
scattering angles are distributed like ${d\sigma(\sqrt{s}) \over d \Omega }$.  
This procedure is fast, precise and easy to control and will replace
in near future all the standard cross section routines. 

In the present case we implemented the measured $KN$ cross sections for the elastic
channels \cite{cam,bar,cha,gi2,dam} and the charge exchange channel with an
angular distribution \cite{gi1}
fitted to the experimental data by use of a neural network as described above.
We also implemented the $\Lambda N$ elastic scattering as isotropic with a total cross
section of 16.4 mb \cite{and}.

\section{Comparison with FOPI data}
These neural networks are embedded in the Quantum Molecular Dynamics (QMD) 
approach to simulate heavy ion reactions on a event by event basis. This
approach simulates the time evolution of all projectile and target nucleons
from their initial separation in projectile and target up to their final
fate being protons, neutrons or part of a cluster. 
The nucleons are represented by coherent state wave functions
which depend on two parameters, the position $r_{i0}$ and the momentum $p_{i0}$.
The Wigner transforms of these coherent states are gaussians in momentum and
coordinate space, respectively. The time evolution of the centroids of the gaussians
$r_{i0}$, $p_{i0}$ is determined by a generalized Ritz variational principle.
The nucleons interact by mutual two and three-body potentials which reduce to a
Skyrme potential in nuclear matter. For details we refer to ref \cite{aic}.

These simulations have been frequently employed to understand the physical
origin of the observed spectra or to predict the observable consequences
of the variation of yet unknown properties of hadronic matter like the
equation of state or in medium properties of particles. 

The kaon are produced with the cross section of Randrup and Ko \cite{cc}
who parametrized the little known cross sections $NN\rightarrow NK\Lambda$
and $NN\rightarrow NK\Sigma$.
Because the rescattering cross section of $\Sigma$'s at the energy of interest
is not known, we apply for the $\Sigma$N collisions the same cross section 
as for the $\Lambda$N collisions. Finally we let decay the $\Sigma^0$'s into
$\Lambda$'s in order to obtain the $\Lambda$ spectrum. 
Due to the very similar mass
of $\Lambda$'s and $\Sigma$'s the error of this approximation is small
as far as the kaons are concerned and in any case smaller than the systematic error
caused by the ignorance of the $np \rightarrow \Lambda$ cross section.
This cross section depends on the particle which is exchanged which may be
a kaon or a pion. For a pion exchange the cross section would be 2.5 time
larger than that of  the pp channel, for a kaon exchange both are the same. 

From the moment of their production the kaons 
move on straight line trajectories as free particles with their
on shell mass. If kaons come closer to a nucleon
than $r=\sqrt{\sigma_{tot}/\pi}$ they rescatter with
an angular distribution of the free scattering. Charge exchange
reactions and elastic scattering  have a relative fraction 
$\sigma_{ch\ ex}/\sigma_{tot}$ and  $\sigma_{el}/\sigma_{tot}$, respectively
with $\sigma_{tot}=\sigma_{ch \ ex}+\sigma_{el}$ . $\Lambda$'s rescatter with
nucleons with a cross section given by \cite{and} and feel the same potential
as the nucleons.

For the comparison we have chosen the reaction Ni(1.93AGeV)+Ni measured by
the FOPI collaboration at GSI \cite{bes}. First, because in this experiments the
rapidity and transverse energy distribution of the kaons has been measured,
second, because the low observed kaon flow brought up the conjecture
that it is caused due to the balance of the scalar and the 
vector potential\cite{gqli}.

We start our comparison with the rapidity distribution of kaons which is
represented in fig.3. We compare the rapidity distribution of kaons
as created and after rescattering with the experimental results. The
experimental points are supplemented with a preliminary experimental data point
at midrapidity obtained by the KAOS collaboration at 1.8 GeV/N for the
same projectile target combination\cite{se}. According to our calculations 
this point has to be multiplied by 1.1 in order to be comparable with
the results at 1.93 GeV .Without
rescattering the results correspond to the distribution obtained by
the perturbative treatment. We observe, first of all,
a surprisingly large influence of rescattering even for systems as
small as Ni + Ni. This confirms calculations of Fang et al \cite{fa}.
That rescattering enlarges the width in rapidity is
expected: Due to the limited available energy the creation of strange
particles is centered
around midrapidity and the width of the created kaons, given by phase space,
is very limited. Rescattering tries to "thermalize" the kaons, hence
the rapidity approaches that of the nucleons which is displayed in fig. 4.
We see that the kaons have finally still a smaller variance
than the nucleons which are far from representing a thermal system. 
Despite we are well above the threshold about 50 \% of the kaons
come from $\Delta$N and only the rest from NN collisions.

In fig. 3 we compare as well our calculation with that of G.Q.Li et al.
\cite{lik,liko}. They have used a relativistic RBUU program and an 
isotropic cross section of 10 mb for KN rescattering \cite{fa}. Both programs
differ in their nucleon and kaon  potentials. Whereas in QMD we use 
for the nuclear potential a nonrelativistic
Skyrme type potential with a soft equation of state
as already in ref.\cite{har}, Li et al. use a relativistic vector and scalar
potential. At energies above 1 GeV/N the optical potential becomes energy
independent and the energy dependence at lower energies becomes only 
important after the kaons have been produced. Hence we have neglected
the energy dependence. This constancy of the  optical potential 
cannot be reproduced in relativistic models which are bound to give (for energy
independent coupling constants) a linear dependence on the kinetic energy.
Beside this difference the Schr\"odinger equivalent 
potential of the relativistic potential used by Li et al. is not far 
away from the Skyrme potential applied here. In addition, above threshold the influence of the
potential becomes less and less important. 
The kinematics is relativistic in  both cases. The major difference between
both approaches is the use of a kaon optical potential by Li et al. 
(consisting of a scalar and a vector part).  In our
calculation the kaons move in between collisions on straight lines. 

It is surprising how little these differences influence the kaon rapidity
distribution. Especially the use of a kaon optical  potential seems not
to have any observable influence on this variable.

The similar broadening of the distribution by rescattering is observed
for the
transverse energy spectra
$\frac{1}{M_{\bot}^2}\frac{d^2N}{dY^{(0)}dM_{\bot}}=Ae^{-M_{\bot}/T_B}$, fig.5,
where we compare the 
different slopes with those measured by the FOPI collaboration.
In both cases this slope is fitted to the high energy tail of the spectra.
Before rescattering the slope is determined by the available
phase space in the production collision. Subsequent collisions increase
the slope $T_B$ of the transverse energy spectra 
by about 70\%, almost independent of the rapidity. We would like to
mention that the increase of the slope is much larger than the
increase of the average transverse root mean square momentum,
which is about 25\%.   

The influence of the rescattering on the observed spectra is
displayed in fig.6 where we show for 3 laboratory angles ( 
$\theta_{lab} = 0^\circ$,\ $\theta_{lab} = 44^\circ$, 
the standard angle for the kaons observed by the KAOS collaboration, and
$\theta_{lab} = 85^\circ$) the spectra with (WR) and without (NR)
rescattering. We see an influence of the
rescattering which is not at all negligible, however less dramatic
as in ref.\cite{fa}, where kaon rescattering at lower energies
in a smaller system has been investigated.

Fig.7 shows the flow of kaons and $\Lambda$'s. We display the 
average in plan transverse momentum as a function of the rapidity.
On the left hand side we have included all the particles on the right
hand side those with $p_t/m > 0.5 $ which corresponds to the acceptance of the
FOPI detector. We see that for kaons before and after rescattering the 
$<p_x>$ is compatible with 0 whereas the $\Lambda$'s seem to show
a finite flow even without rescattering. 

Kaons and $\Lambda$'s are
produced according to phase space and hence isotropically in the NN center
of mass system. Hence only a finite average transverse velocity of the NN
center of mass system can be the origin of a finite average transverse
momentum of the particles when they get produced. This transverse velocity
$<v_x(y)> = <{p_x(y)\over E}>$ as a function of rapidity is displayed in fig. 8.
However, a common center of mass velocity has not the same consequences for 
$\Lambda$'s and kaon's. The root mean square rapidity of
kaons in the NN center of mass system is 70\% larger as that for the
$\Lambda$'s. Hence the kaons average a finite transverse source
velocity over a larger rapidity range (the variance is about .28 $y_0$).
This averaging  leads - besides close to midrapidity - to a smaller 
transverse velocity as compared to that of the $\Lambda$'s.
This is seen in fig. 8 as well. 

The KN cross section is not only much smaller as the $N\Lambda$ cross section
but as well forward peaked in the momentum range of interest. Thus if the
kaon, produced around midrapidity, hits a nucleon of the in streaming matter it
changes its direction only little as compared to the $\Lambda$ which
is assumed to have an isotropic cross section. As a consequence, a collective
transverse velocity is much less communicated to the kaons. Due to the geometry
the forward
peaked cross section makes it easier for the kaons to escape from the system.
This, as well as the larger velocity of the kaons results in a small
number of KN collisions. Indeed 
the kaons suffer fewer collisions as compared to the $\Lambda$'s as expected 
from the cross section ratios. We see in fig. 7 that rescattering does not give 
the kaons a directed transverse momentum whereas for the $\Lambda$'s this
is observed. Another consequence of this dynamics is the small increase
of the variance in rapidity of the kaons (12\%) due to rescattering
as compared to that of the $\Lambda$ 's (39\%) which will be discussed
later.
We see that the velocity of the final $\Lambda$'s
follows that of the nucleons except for the large rapidity were
the spectators contribute. Hence rescattering increases the transverse
flow of the $\Lambda$'s to that extend that it agrees with that of 
the protons as seen in fig.7. 

Hence we cannot confirm the conjecture that the kaon flow may be of
use to measure the scalar and vector part of the kaon potential \cite{liko}.
No kaon potential as in the present investigation yields already
the observed flow. Of course there is a kaon potential but the only information
one may extract from the observable flow is that the combination
of vector and scalar potential has to have no influence on that observable.

In view of former studies this is a very expected result. Kaons are
predominately produced in central collisions where the collective 
flow is small \cite{har}. In addition they are produced in the high
density zone. Nucleons from the high density zone show a smaller flow
than the average over all nucleons \cite{jae} because the flow is caused by 
the potential gradient which diverts the nucleons from the high density zone.

Fig. 9 displays the distribution of nuclear densities at the positions
were the kaons are created and at the point of their last rescattering. 
We see that the average density at the point of creation is well above 
normal nuclear matter density. 79\% of the kaons are produced at $\rho > 
\rho_0$. Even further interactions do not spoil the ability of the
kaons to transfer informations of the high density zone to the detectors.
60.5\% of the kaons did not suffer rescattering or had their last collision
at a density larger then normal nuclear matter density. 

Finally, in fig. 10 we display the rapidity distribution of the $\Lambda$'s before (dashed)
and after (full line) rescattering. The absolute value here is even more 
plagued by the
little knowledge of the exclusive cross section $NN\rightarrow NK\Sigma$
resp. $NN\rightarrow NK\Lambda$
than that for the kaons, especially in cases where a neutron is
in the entrance channel. Measurements do not exist there and theoretical
calculations   \cite{lag,lik} predict quite different results depending 
on the type of the  exchanged meson (kaon or pion).
Preliminary experimental results of the FOPI collaboration 
come close to our rapidity distribution, for the absolute values one
has, however, await the final analysis.

\section{Conclusion}
In conclusion we have shown, that despite of the small cross section 
the rescattering influences the strange particle observables 
even for small systems. It widens the 
energy distribution in transverse and in longitudinal direction.
A realistic treatment with takes into account all presently known
differential cross sections reproduces the kaon observables for
the case investigated despite of the fact that no kaon potential
has been employed. This leaves little room for the conjecture
that $K^+$ observables may yield information on the kaon potential
in nuclear matter in a unique way. The experimental results 
are compatible with calculations which show that
the $K^+$ properties change only little in nuclear matter. The situation
is probably quite different for $K^-$. This is presently under investigation.

We would like to thank Dr.D.Best for his preparation of some of the figures
Drs G.Q. Li and C.M. Ko for the permission to use their results and Drs. 
D. Best, N. Herrmann, C.M. Ko, G.Q. Li and J. Ritman for discussions.

\newpage
\bf{FIGURE CAPTIONS}\\
\\
\noindent
\normalsize
Fig.1: Schematic Neural Network used for $KN$ scattering.\\
\\
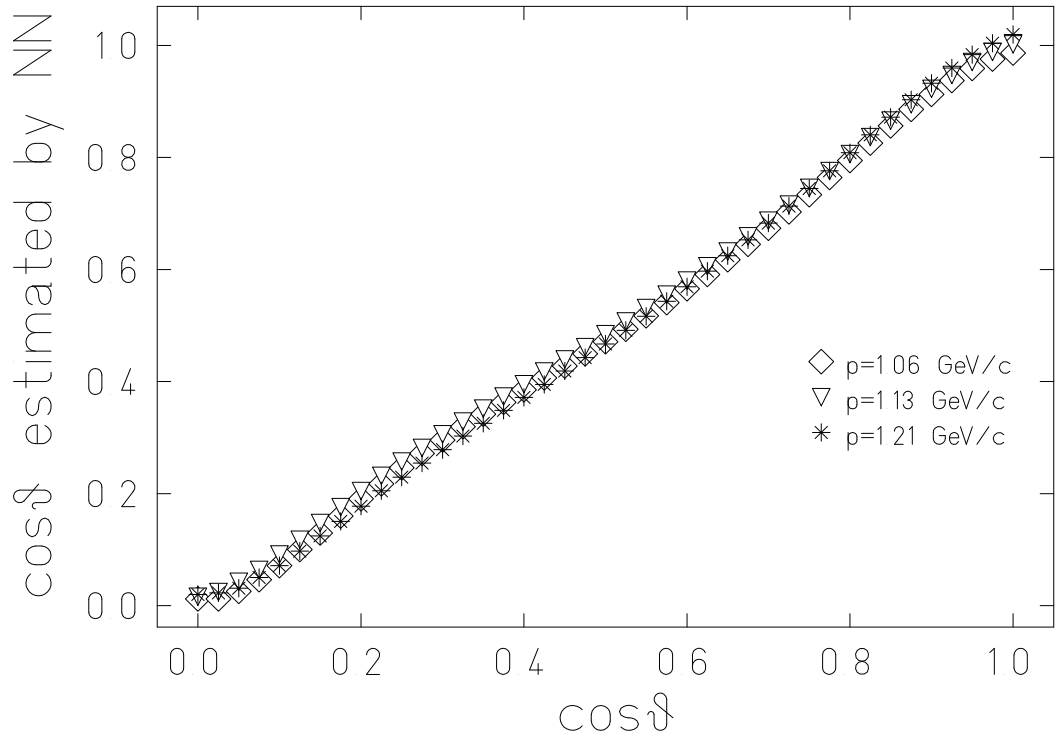
Fig.2: Output of the Neural Network for $KN$ charge exchange scattering with
$p_{lab}=1.06, 1.13, 1.21 GeV/c$ as compared to the input.\\
\\
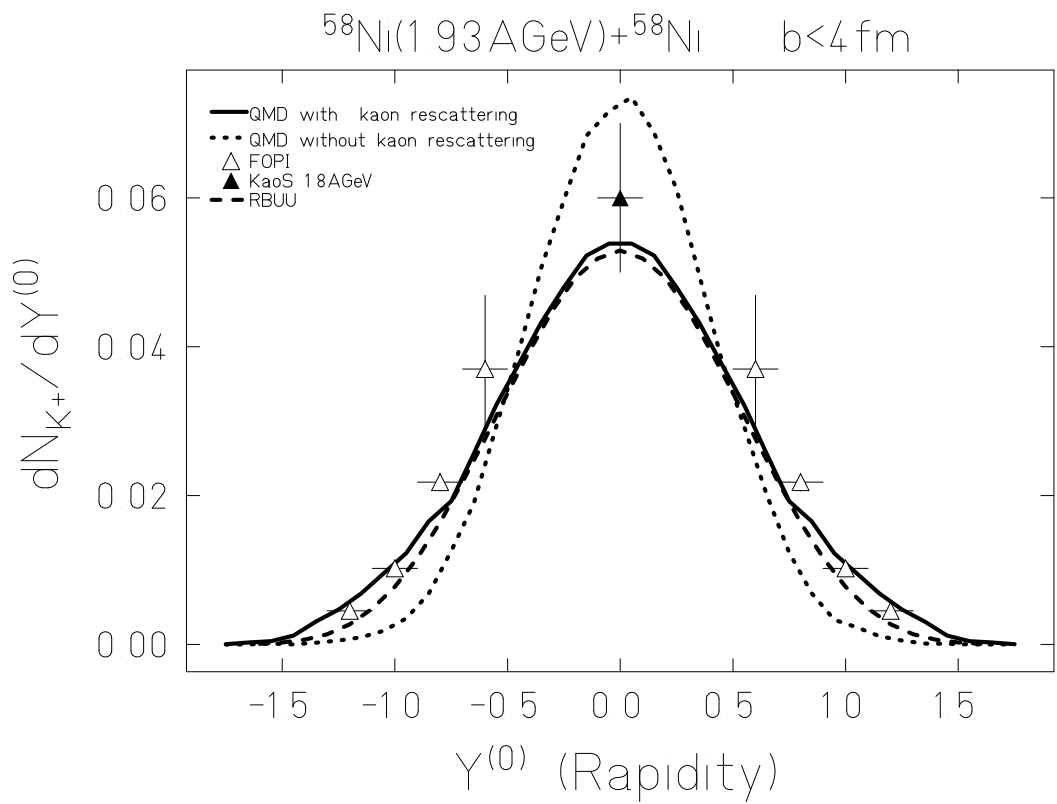
Fig.3: Rapidity distribution of kaons with and without rescattering as
compared to FOPI data \cite{bes} and the calculation of Li et al \cite{gqli}.\\
\\
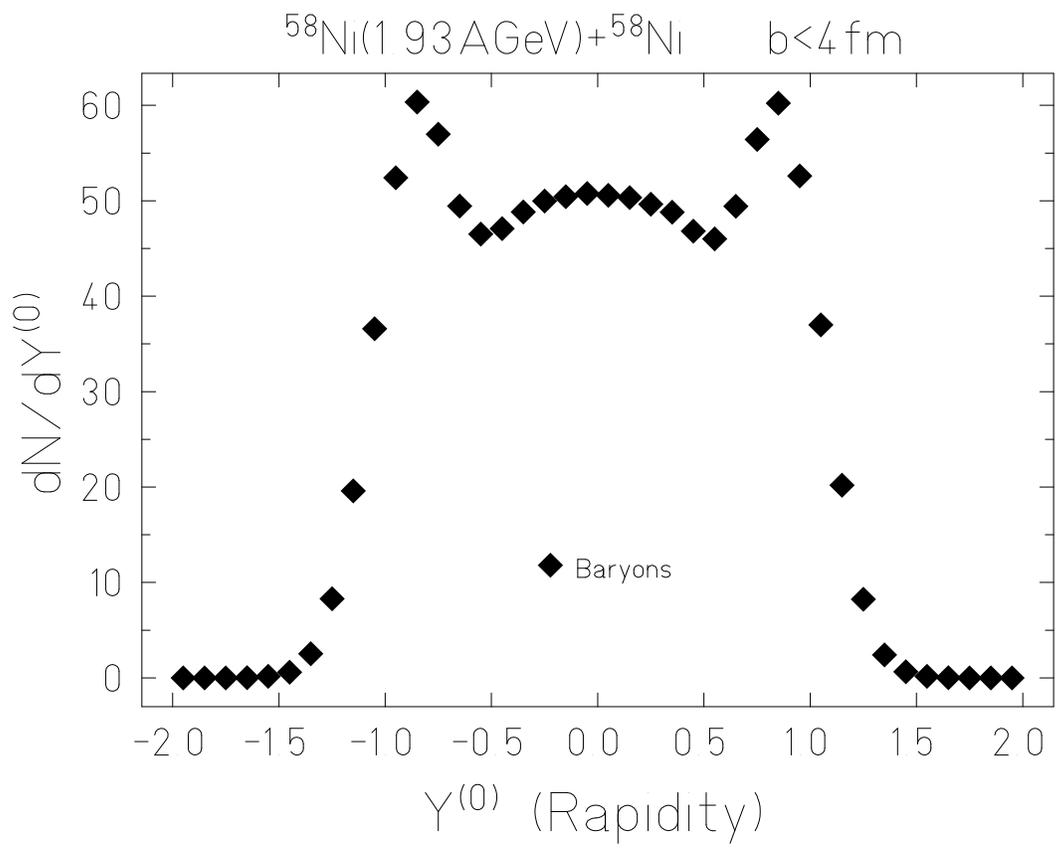
Fig.4: Rapidity distribution of baryons.\\
\\
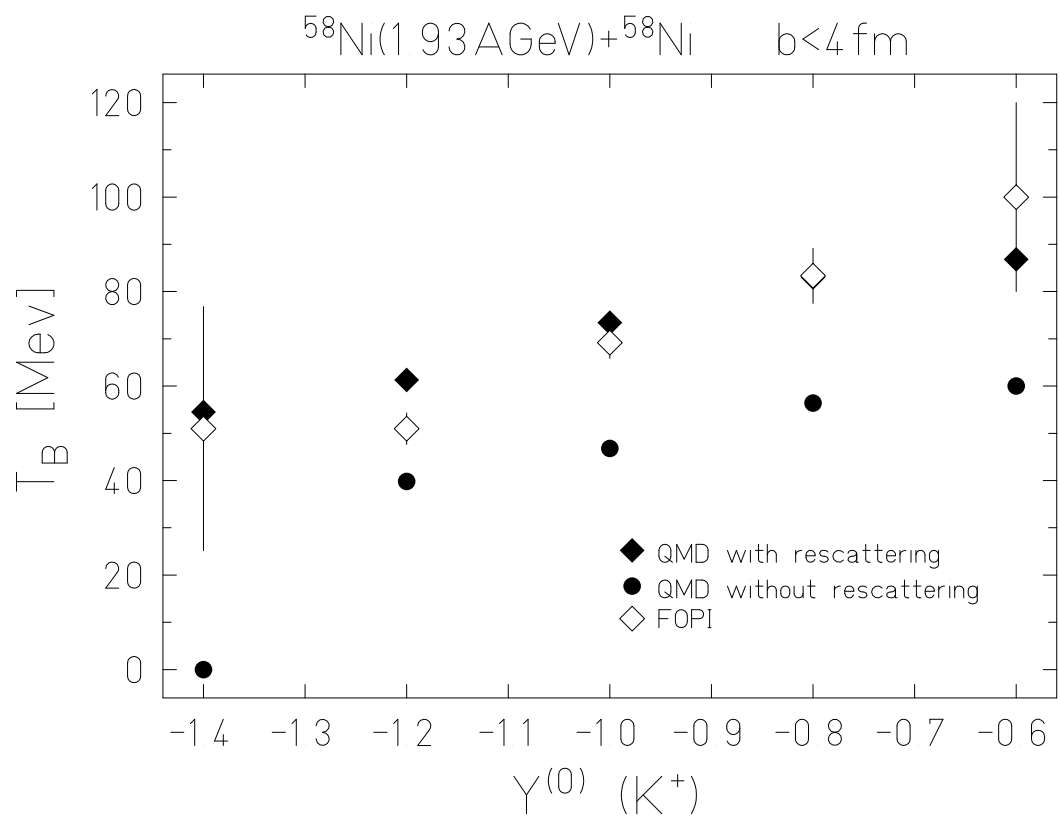
Fig.5: Slopes $T_B$ of the spectra
$\frac{1}{M_{\bot}^2}\frac{d^2N}{dY^{(0)}dM_{\bot}}=Ae^{-M_{\bot}/T_B}$ as a function of the rapidity of kaons with and without rescattering
as compared to FOPI data \cite{bes}.\\
\\
Fig.6: $\frac{d^2\sigma}{dp_{lab}d\Omega}$ of kaons as a function of  $p_{lab}$ for
$\theta_{lab}=0^\circ,44^\circ,85^\circ$.\\
\\
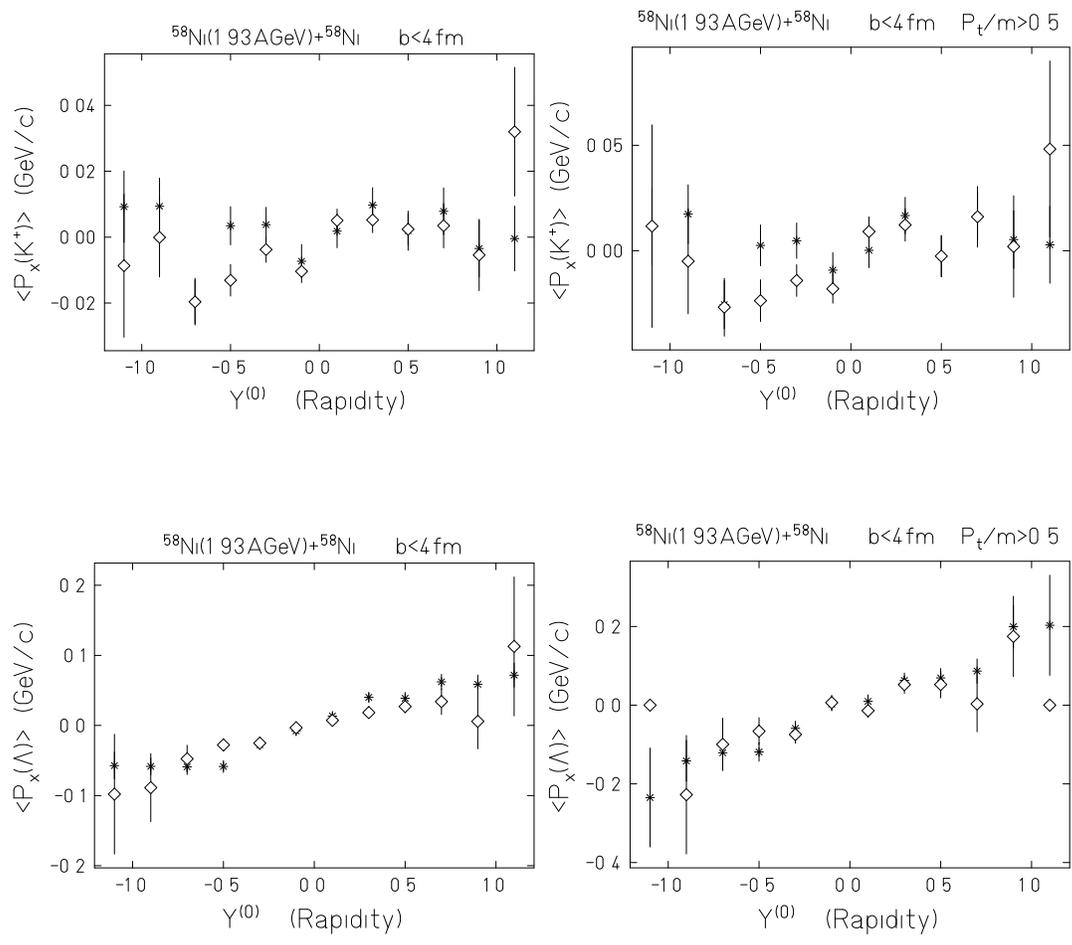
Fig.7: $<P_x(y)>$ for kaons and $\Lambda$'s with and without rescattering. On
the left hand side we include all particles, on the right hand side we applied
the acceptance filter of FOPI.\\
\\
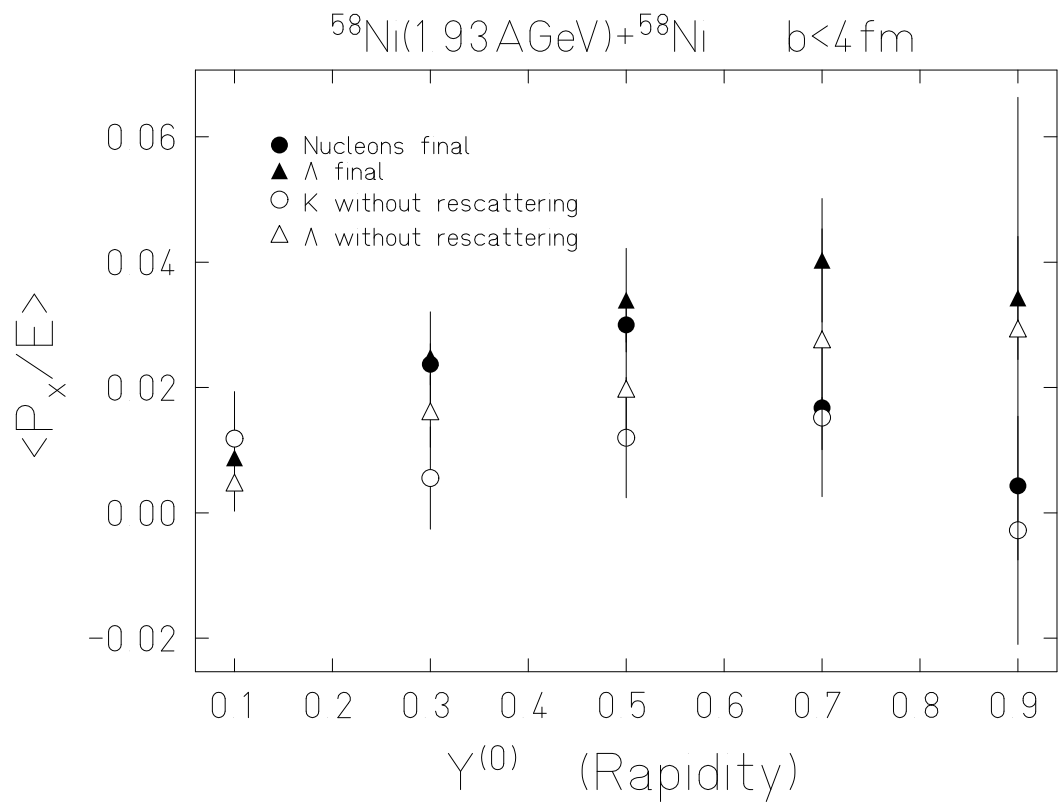
Fig.8: Velocity on $x$ direction as a function of the rapidity for kaons,
 $\Lambda$'s and nucleons.\\
\\
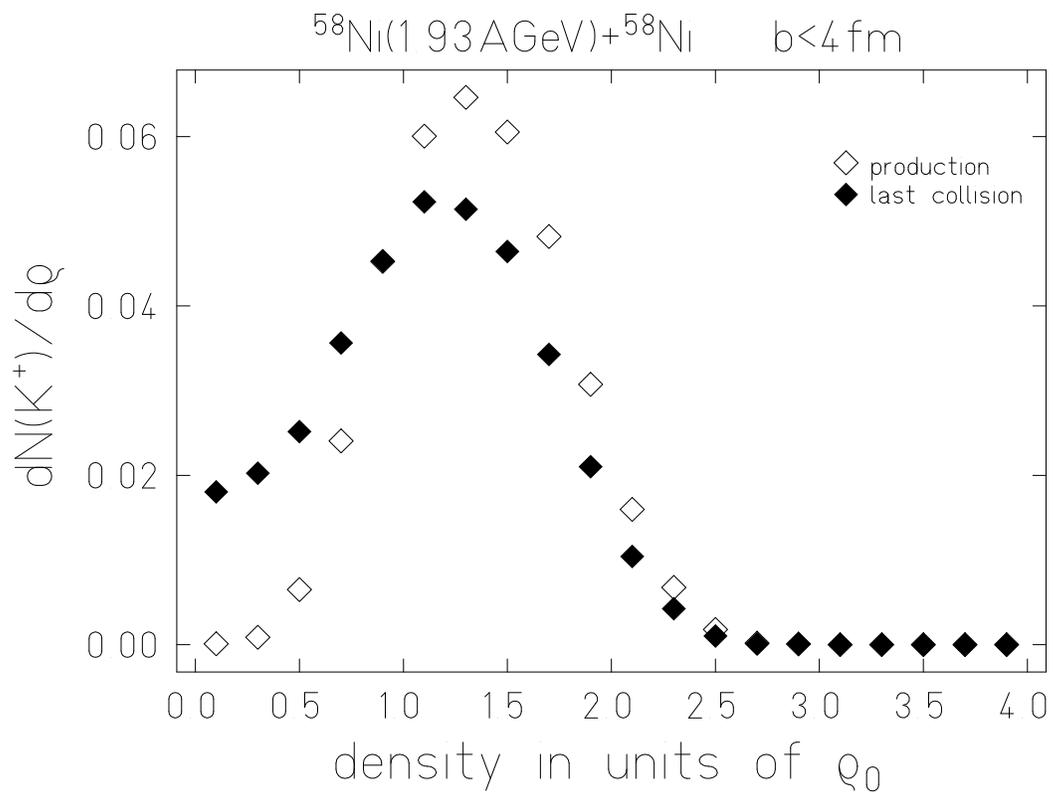
Fig.9: Density distribution of kaons at the production points and at the points
of the last collision.\\
\\
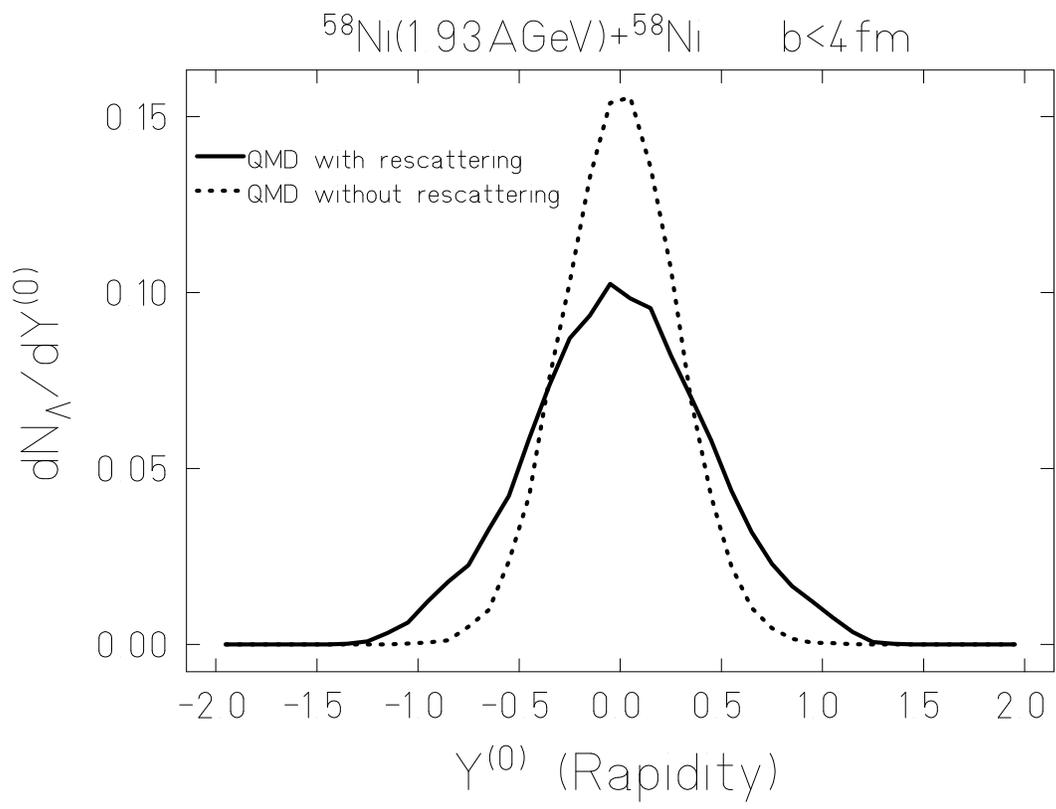
Fig.10: Rapidity distribution for $\Lambda$'s with and without rescattering.\\
\\

\newpage
\begin{thebibliography}{99}
\bibitem{aik} J. Aichelin and C.M. Ko Phys. Rev. Lett. 55 (1985) 2661
\bibitem{har} C.Hartnack et al., Nucl. Phys. A580 (1994) 643
\bibitem{sch} J. Schaffner et al, Phys. Lett B334 (1994) 268
\bibitem{hog} W.J. Hogan, Phys. Rev. 166 (1968) 1472
\bibitem{lag} J.M. Laget, Phys. Lett. B 259 (1991) 24
\bibitem{ran} J.Randrup, Phys.Lett. B99 (1981) 9
\bibitem{fa} X.S. Fang , C.M. Ko and  Y.M. Zheng Nucl. Phys. A556 (1993) 499
\bibitem{dav} C.David et al., Phys. Rev. C51 (1995) 1453
\bibitem{cam} W.Cameron et al., Nucl. Phys. B78 (1974) 93
\bibitem{bar} P.C.Barber et al., Nucl. Phys. B61 (1973) 125
\bibitem{cha} B.J.Charles et al., Nucl. Phys. B131 (1977) 7
\bibitem{gi2} G.Giacomelli et al., Nucl. Phys. B56 (1973) 346
\bibitem{dam} C.J.S Damerell et al., Nucl. Phys. B94 (1975) 374
\bibitem{gi1} G.Giacomelli et al., Nucl. Phys. B42 (1972) 437
\bibitem{and} Anderson et al., Phys. Rev. D11 (1975) 473
\bibitem{aic} J. Aichelin, Phys. Rep. 202 (1991) 233
%\bibitem{wd} Compilation of cross sections CERN - HERA 84 -01
\bibitem{cc}J. Randrup and C.M. Ko, Nucl. Phys. A 343 (1980) 519 and A 411
\bibitem{bes} D.Best for the FOPI collaboration, Proc. on the $12^{th}$ Winter
Workshop on Nuclear Dynamics, Snowbird, Utah, Feb.1996, ed.W.Bauer
\bibitem{gqli} G.Q. Li and C.M. Ko, private communication
\bibitem{se} P. Senger, to be published in Heavy Ion Phys. (1996)
and private communication 
\bibitem{lik} G.Q. Li and C.M. Ko, Nucl. Phys. A 594 (1995) 460
\bibitem{liko} G.Q. Li, C.M. Ko and B.A. Li, Phys. Rev. Lett 74. (1995) 235
\bibitem{jae} J. Jaenicke and J. Aichelin, Nucl. Phys. A 547 (1992) 542
\end{thebibliography}
\end{document}